\documentclass{PoS}

\usepackage{amssymb,amsmath,slashed,slashbox,multirow,amsfonts,latexsym,dsfont,bm}

\newcommand{\be}{\begin{equation}}
\newcommand{\ee}{\end{equation}}
\newcommand{\ud}{\mathrm{d}}
\newcommand{\uTr}{\mathrm{Tr}}

\newcommand{\uvec}[1]{\boldsymbol{#1}}

\newlength\savedwidth

\newcommand\whline{\noalign{\global\savedwidth\arrayrulewidth
\global\arrayrulewidth 1pt}%
\hline
\noalign{\global\arrayrulewidth\savedwidth}}

\title{Transverse phase space and its multipole decomposition}

\ShortTitle{Transverse phase space and its multipole decomposition}

\author{\speaker{C\'edric Lorc\'e}\\
       Centre de Physique Th\'eorique, \'Ecole polytechnique, CNRS, Universit\'e Paris-Saclay, F-91128 Palaiseau, France\\
       E-mail: \email{cedric.lorce@polytechnique.edu}}

\author{Barbara Pasquini\\
Dipartimento di Fisica, Universit\`a degli Studi di Pavia, Pavia, Italy\\
Istituto Nazionale di Fisica Nucleare, Sezione di Pavia, Pavia, Italy\\
        E-mail: \email{pasquini@pv.infn.it}}

\abstract{Relativistic phase space distributions are very interesting objects as they allow one to gather the information extracted from various types of experiments into a single coherent picture. Focusing on the four-dimensional transverse phase space, we identified all the possible angular correlations providing at the same time a clear physical interpretation of all the leading-twist generalized and transverse-momentum dependent parton distributions. We also developed a convenient representation of this four-dimensional space.}

\FullConference{QCD Evolution 2016\\
		May 30-June 03, 2016\\
		National Institute for Subatomic Physics (Nikhef), Amsterdam}

\begin{document}

\section{Introduction}

The concept of phase-space distribution can be carried over to the context of Quantum Field Theory. A six-dimensional version has been introduced in Refs.~\cite{Ji:2003ak,Belitsky:2003nz} but   is valid only for infinitely massive targets in order to get rid of relativistic corrections. For finite target mass, a five-dimensional phase-space distribution free of relativistic corrections can however be introduced within the light-front formalism~\cite{Lorce:2011kd} and appears to be the Fourier transform of Generalized Transverse-Momentum dependent Distributions (GTMDs)~\cite{Meissner:2009ww,Lorce:2011dv,Lorce:2013pza}. The latter are in some sense the \emph{mother distributions} of Generalized Parton Distributions (GPDs) and Transverse-Momentum dependent Distributions (TMDs), and provide a natural access to the parton orbital angular momentum (OAM)~\cite{Lorce:2011kd,Hatta:2011ku,Lorce:2011ni,Kanazawa:2014nha,Liu:2015xha}. Currently, the best hope to access directly these GTMDs is in the low-$x$ regime~\cite{Martin:1999wb,Khoze:2000cy,Martin:2001ms,Albrow:2008pn,Martin:2009ku,Hatta:2016dxp}.

At leading twist, there are 32 quark phase-space distributions, half of them being associated to naive $\mathsf T$-odd GTMDs and hence encoding initial and/or final-state interactions. A detailed study of these distributions has been presented in Refs.~\cite{Lorce:2011kd,Lorce:2015sqe}. Here we will focus on the multipole structure of the transverse phase space obtained by integrating the five-dimensional phase-space distributions over the parton longitudinal momentum. This allows us to identify the various possible angular correlations and to determine how they are encoded in the GPDs and TMDs, easier to access experimentally.

The plan of the paper is as follows.
 In Sec.~\ref{section:1} we define the Wigner distribution as a Fourier transform of the GTMD correlator to the impact-parameter
space, and we present its properties under parity and time-reversal transformations.
 In Sec.~\ref{sec:multipole}, we decompose the Wigner functions in terms of basic multipoles in the transverse phase space and coefficient functions, and we summarize all the possible angular correlations encoded in these phase-space distributions. 
In Sec.~\ref{section:3} we sketch a new way for depicting the transverse phase space, which allows one to visualize the multipole structures simultaneously in both the transverse-momentum and transverse-position spaces.
 In Sec.~\ref{section:4} we discuss the results for the lowest multipole structure and make the connection with both GPDs and TMDs. A specific relativistic light-front constituent quark model~\cite{Lorce:2011dv} has been used to check the generic decomposition and illustrate particular multipole structures. Finally, we gather our conclusions in Sec.~\ref{section:5}.

\section{Polarized relativistic phase-space distributions}
\label{section:1}

We adopt the light-front formalism where the components of a four-vector $a^\mu$ are given by $[a^+,a^-,\uvec a_T]$ with $a^\pm=\tfrac{1}{\sqrt{2}}(a^0\pm a^3)$. The quark GTMD correlator is then defined as~\cite{Meissner:2009ww,Lorce:2013pza}
\begin{equation}\label{GTMDcorr-def}
W^{ab}_{\Lambda'\Lambda}\equiv\int\ud k^-\int\frac{\ud^4z}{(2\pi)^4}\,e^{ik\cdot z}\,\langle P+\tfrac{\Delta}{2},\Lambda'|\overline\psi_b(-\tfrac{z}{2})\,\mathcal W\,\psi_a(\tfrac{z}{2})|P-\tfrac{\Delta}{2},\Lambda\rangle.
\end{equation}
$\mathcal W$ is a Wilson line ensuring color gauge invariance, $k$ is the quark average four-momentum, and $|p,\Lambda\rangle$ is the spin-$1/2$ target state with four-momentum $p$ and light-front helicity $\Lambda$. A more consistent definition of the GTMDs should in principle also include a soft factor contribution~\cite{Echevarria:2016mrc}, but the latter has no real impact on the following discussions and can therefore be omitted. We choose to work in the symmetric frame defined by $P^\mu=\tfrac{p'^\mu+p^\mu}{2}=[P^+, P^-,\uvec 0_T]$. At leading twist, one can interpret
\begin{equation}\label{GTMDcorr}
W_{\vec S\vec S^q}=\tfrac{1}{8}\sum_{\Lambda',\Lambda}(\mathds 1+\vec S\cdot\vec \sigma)_{\Lambda\Lambda'}\,\uTr[W_{\Lambda'\Lambda}\Gamma_{\vec S^q}]
\end{equation}
with $\Gamma_{\vec S^q}=\gamma^++S^q_L\,\gamma^+\gamma_5+S^{qj}_T\,i\sigma^{j+}_T\gamma_5$,
as the GTMD correlator describing the distribution of quarks with polarization $\vec S^q$ inside a target with polarization $\vec S$~\cite{Lorce:2011zta}.

The corresponding phase-space distribution is obtained by Fourier transform~\cite{Lorce:2011kd}
\begin{equation}
\rho_{\vec S\vec S^q}(x,\uvec k_T,\uvec b_T;\hat P,\eta)=\int\frac{\ud^2\Delta_T}{(2\pi)^2}\,e^{-i\uvec\Delta_T\cdot\uvec b_T}\,W_{\vec S\vec S^q}(P,k,\Delta)\big|_{\Delta^+=0},
\end{equation}
and can be interpreted as giving the quasi-probability of finding a quark with polarization $\vec S^q$, transverse position $\uvec b_T$ and light-front momentum $(xP^+,\uvec k_T)$ inside a spin-$1/2$ target with polarization $\vec S$~\cite{Lorce:2011kd}. The direction of the average target momentum is given by $\hat P=\vec P/|\vec P|$ and the parameter $\eta$ indicates whether $\mathcal W$ goes to $+\infty^-$ or $-\infty^-$. Because of the hermiticity property of the GTMD correlator~\eqref{GTMDcorr}, this phase-space distribution is always real-valued. Moreover, under parity and time reversal, it behaves as
\begin{equation}
\begin{aligned}
\rho_{\vec S\vec S^q}(x,\uvec k_T,\uvec b_T;\hat P,\eta)&\stackrel{\mathsf P}{\mapsto}\rho_{\vec S\vec S^q}(x,-\uvec k_T,-\uvec b_T;-\hat P,\eta),\\
&\stackrel{\mathsf T}{\mapsto}\rho_{-\vec S-\vec S^q}(x,-\uvec k_T,\uvec b_T;-\hat P,-\eta).
\end{aligned}
\end{equation}
There are 16 independent polarization configurations~\cite{Lorce:2011kd,Lorce:2013pza} corresponding to 16 independent linear combinations of GTMDs~\cite{Meissner:2009ww,Lorce:2013pza}. Each polarization configuration can further be decomposed into naive $\mathsf T$-even and $\mathsf T$-odd contributions
\begin{equation}
\rho_{\vec S\vec S^q}=\rho^e_{\vec S\vec S^q}+\rho^o_{\vec S\vec S^q},
\end{equation}
where 
\begin{equation}
\begin{aligned}
\rho^e_{\vec S\vec S^q}(x,\uvec k_T,\uvec b_T;\hat P,\eta)&=+\rho^e_{-\vec S-\vec S^q}(x,-\uvec k_T,\uvec b_T;-\hat P,\eta),\\
\rho^o_{\vec S\vec S^q}(x,\uvec k_T,\uvec b_T;\hat P,\eta)&=-\rho^o_{-\vec S-\vec S^q}(x,-\uvec k_T,\uvec b_T;-\hat P,\eta).
\end{aligned}
\end{equation}
In some sense, $\rho^e$ describes the \emph{intrinsic} distribution of quarks inside the target, whereas $\rho^o$ describes how \emph{extrinsic} initial and/or final-state interactions modify this distribution.

\section{Multipole decomposition}
\label{sec:multipole}

\begin{figure}[t!]
	\centering
		\includegraphics[width=\textwidth]{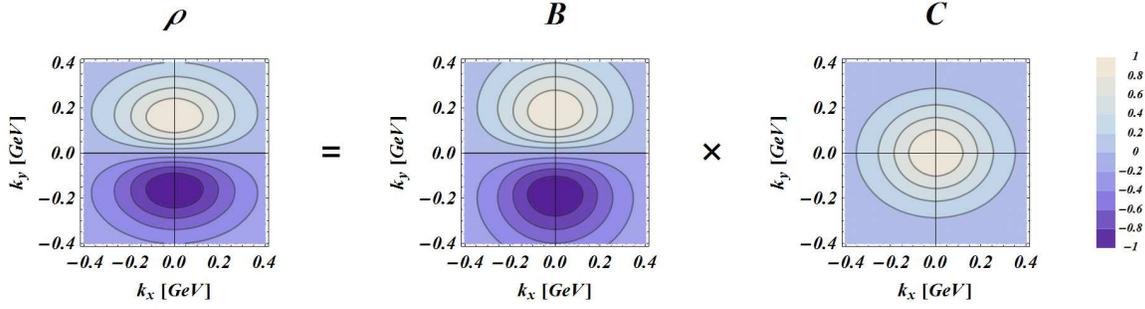}
\caption{\footnotesize{Simple illustration of the decomposition at fixed $x$ and $\uvec b_T$. The phase-space distribution $\rho$ can be written as a product of a basic multipole $B$ (here a dipole in $\uvec k_T$-space) with an oval-shaped coefficient function $C$.}}
		\label{Illustration}
\end{figure}

The relativistic phase-space distribution is linear in $\vec S$ and $\vec S^q$ 
\begin{equation}
\begin{aligned}
\rho_{\vec S\vec S^q}&=\rho_{UU}+S_L\,\rho_{LU}+S^q_L\,\rho_{UL}+S_LS^q_L\,\rho_{LL}\\
&\phantom{=}+S^i_T\,(\rho_{T^iU}+S^q_L\,\rho_{T^iL})+S^{qi}_T\,(\rho_{UT^i}+S_L\,\rho_{LT^i})+S^i_TS^{qj}_T\,\rho_{T^iT^j}.
\end{aligned}
\end{equation}
Each component $\rho_X$ with $X= UU, LU, UL, \cdots$ can further be decomposed into multipoles in both $\uvec k_T$ and $\uvec b_T$ spaces
\begin{align}
\rho_{X}(x,\uvec k_T,\uvec b_T;\hat P,\eta)&=\sum_{m_k,m_b} \rho^{(m_k,m_b)}_{X}(x,\uvec k_T,\uvec b_T;\hat P,\eta),\\
\rho^{(m_k,m_b)}_{X}(x,\uvec k_T,\uvec b_T;\hat P,\eta)&=B^{(m_k,m_b)}_{X}(\hat k_T,\hat b_T;\hat P,\eta)\,C^{(m_k,m_b)}_{X}[x,\uvec k^2_T,(\uvec k_T\cdot\uvec b_T)^2,\uvec b^2_T],\label{Decomposition}
\end{align}
with $B^{(m_k,m_b)}_{X}$ the basic multipoles constrained by parity and time-reversal and $C^{(m_k,m_b)}_{X}$ the coefficient functions which depend on $\mathsf P$ and $\mathsf T$-invariant variables only. The couple of integers $(m_k,m_b)$ gives the order of the basic multipole in $\uvec k_T$ and $\uvec b_T$ spaces. Fig.~\ref{Illustration} gives an illustration of the decomposition of a phase-space density into basic multipole and coefficient function.

Note that only the multipoles with $m_b=0$ survive integration over $\uvec b_T$ and reduce to TMD amplitudes. Similarly, only the multipoles with $m_k=0$ survive integration over $\uvec k_T$. The naive $\mathsf T$-even ones correspond to impact-parameter distributions, i.e. Fourier transforms of GPD amplitudes~\cite{Lorce:2011dv,Diehl:2005jf}. Interestingly, the naive $\mathsf T$-odd ones correspond to \emph{new} contributions appearing also in the general parametrization of the light-front energy-momentum tensor~\cite{Lorce:2015lna}.

\begin{table}[t!]
\begin{center}
\begin{tabular}{@{\quad\!}c@{\quad}|@{\quad}c@{\quad}c@{\quad}c@{\quad}c@{\quad\!}}\whline
$\rho_X$&$U$&$L$&$T_x$&$T_y$\\
\hline
$U$&$\langle 1\rangle$&$\langle S^q_L\ell^q_L\rangle$&$\langle S^q_x\ell^q_x\rangle$&$\langle S^q_y\ell^q_y\rangle$\\
$L$&$\langle S_L\ell^q_L\rangle$&$\langle S_LS^q_L\rangle$&$\langle S_L\ell^q_LS^q_x\ell^q_x\rangle$&$\langle S_L\ell^q_LS^q_y\ell^q_y\rangle$\\
$T_x$&$\langle S_x\ell^q_x\rangle$&$\langle S_x\ell^q_xS^q_L\ell^q_L\rangle$&$\langle S_xS^q_x\rangle$& $\langle S_x\ell^q_xS^q_y\ell^q_y\rangle$\\
$T_y$&$\langle S_y\ell^q_y\rangle$&$\langle S_y\ell^q_yS^q_L\ell^q_L\rangle$& $\langle S_y\ell^q_yS^q_x\ell^q_x\rangle$&$\langle S_yS^q_y\rangle$\\
\whline
\end{tabular}
\caption{\footnotesize{Correlations between target polarization ($S_L,\uvec S_T$), quark polarization ($S^q_L,\uvec S^q_T$) and quark OAM ($\ell^q_L,\uvec{\ell}^q_T$) encoded in the different phase-space distributions $\rho_X$.}}\label{angcorr}
\end{center}
\end{table}

It turns out that the contributions $\rho_X$ can be understood as encoding all the possible angular momentum correlations, see Table~\ref{angcorr}. Note that $\vec \ell^q$ refers to the \emph{canonical} quark OAM, since it is defined in terms of the canonical quark momentum $\vec k$~\cite{Lorce:2012ce}. The relation with the various angular correlations becomes particularly transparent once one sees the five-dimensional relativistic phase-space distributions as six-dimensional distributions integrated over the quark average longitudinal position $b_L=\vec b\cdot\hat P$~\cite{Lorce:2015sqe}
\begin{equation}
\rho_X(x,\uvec k_T,\uvec b_T;\hat P,\eta)=\int\ud b_L\,\rho_X(\vec k,\vec b;\hat P,\eta).
\end{equation}
Working at the level of phase-space distributions gives us much more insight about the physics encoded in the various GPDs and TMDs because integrations over $\uvec b_T$ and $\uvec k_T$ usually hide the precise form of the angular correlation being probed.

\section{Representation of transverse phase space}
\label{section:3}

\begin{figure}[t!]
	\centering
		\includegraphics[width=.45\textwidth]{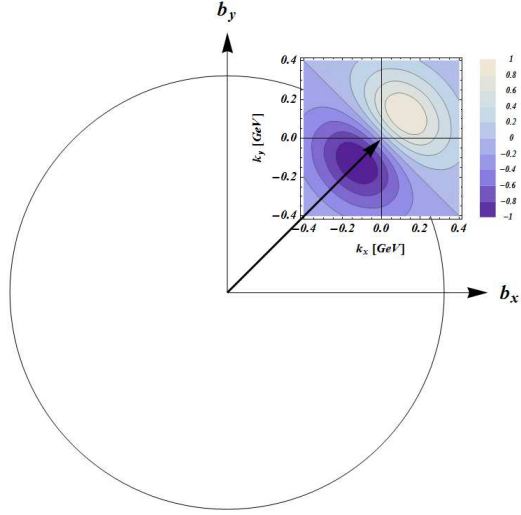}
\caption{\footnotesize{Representation of the transverse phase space. The circle represents the points in impact-parameter space at a fixed distance $|\uvec b_T|$ from the center of the target. To each point on this circle is associated a distribution in transverse-momentum space. See text for more details.}}
		\label{Tphasespace}
\end{figure}

Since we are essentially interested in the traverse phase space $(\uvec k_T,\uvec b_T)$, we reduce the number of variables by integrating the phase-space distributions $\rho_X$ over $x$ and discretizing the polar component of $\uvec b_T$. The resulting transverse phase-space distributions are then represented as sets of distributions in $\uvec k_T$-space 
\begin{equation}
\rho_X(\uvec k_T|\,\uvec b_T)=\int\ud x\,\rho_X(x,\uvec k_T,\uvec b_T;\hat P=\vec e_z,\eta=+1)\big|_{\uvec b_T\text{ fixed}}
\end{equation}
with the origin of axes lying on circles of radius $|\uvec b_T|$ at polar angle $\phi_b$ in impact-parameter space, see Fig.~\ref{Tphasespace}. In this way, one can see how the transverse momentum is distributed at some point in the impact-parameter space. 

This representation of transverse phase space has the advantage of making the multipole structure in both $\uvec k_T$ and $\uvec b_T$ spaces particularly clear. For example, the basic multipole $B^{(m_k,m_b)}_X$ will be represented by a $m_k$-pole in transverse-momentum space at any transverse position $\uvec b_T$, with the orientation determined by $m_b$ and $\phi_b=\arg\hat b_T$. In the following, we chose to represent only eight points in impact-parameter space lying on a circle with radius $|\uvec b_T|=0.4$ fm. Also, for a better legibility, the $\uvec k_T$-distributions are normalized to the absolute maximal value over the whole circle in impact-parameter space
\begin{equation}
\max_{|\uvec b_T|=0.4\text{ fm}}|\rho_X(\uvec k_T|\uvec b_T)|=1.
\end{equation}

\section{Discussion}
\label{section:4}

The results presented here are based on the light-front constituent quark model (LFCQM)~\cite{Lorce:2011dv} for up quarks. Light and dark regions correspond to positive and negative domains of the transverse phase-space distributions, respectively. Here we will focus on a couple of multipole structures only. The interested reader will find the complete discussion in Ref.~\cite{Lorce:2015lna}.

\subsection{$(0,0)$ multipole}

\begin{figure}[t!]
\centerline{\includegraphics[width=7cm]{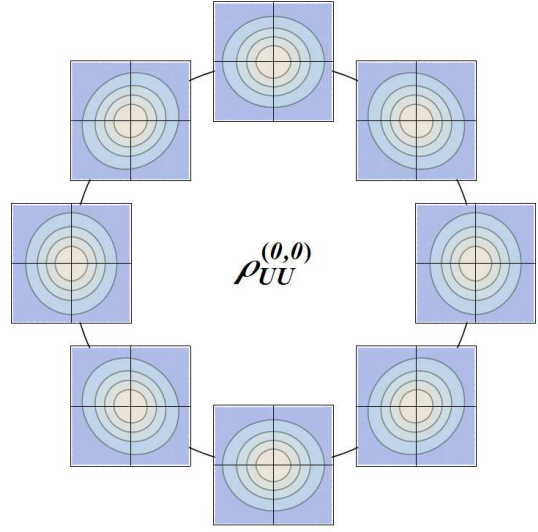}}
\vspace*{8pt}
\caption{The $(0,0)$ multipole appearing in e.g. $\rho^e_{UU}$. See text for more details. \label{fig1}}
\end{figure}

The simplest multipole is naturally the one with $m_k=m_b=0$. It appears in $\rho^e_X$ with $X=UU,LL,TT$ associated to the respective spin structures $1$, $S_LS^q_L$ and $(\uvec S_T\cdot\uvec S^q_T)$. These spin-spin correlations survive integration over $\uvec k_T$ and $\uvec b_T$; they are respectively related to $(H,\tilde H,H_T)$ in the GPD sector and to $(f_1,g_1,h_1)$ in the TMD sector. Contrary to these GPDs and TMDs, $\rho^e_X$ is not circularly symmetric, see Fig.~\ref{fig1}. The reason is that $\rho^e_X$ also contains information about the \emph{correlation} between $\uvec k_T$ and $\uvec b_T$ (encoded in the coefficient functions $C^e_X$ through the $(\hat b_T\cdot\hat k_T)^2$ dependence) which is lost under integration over $\uvec k_T$ or $\uvec b_T$~\cite{Lorce:2011kd}. 

\subsection{$(0,1)$ and $(1,0)$ multipoles}

The first non-trivial multipoles are the ones with $m_k+m_b=1$. 

\begin{figure}[t!]
\centerline{\includegraphics[width=7cm]{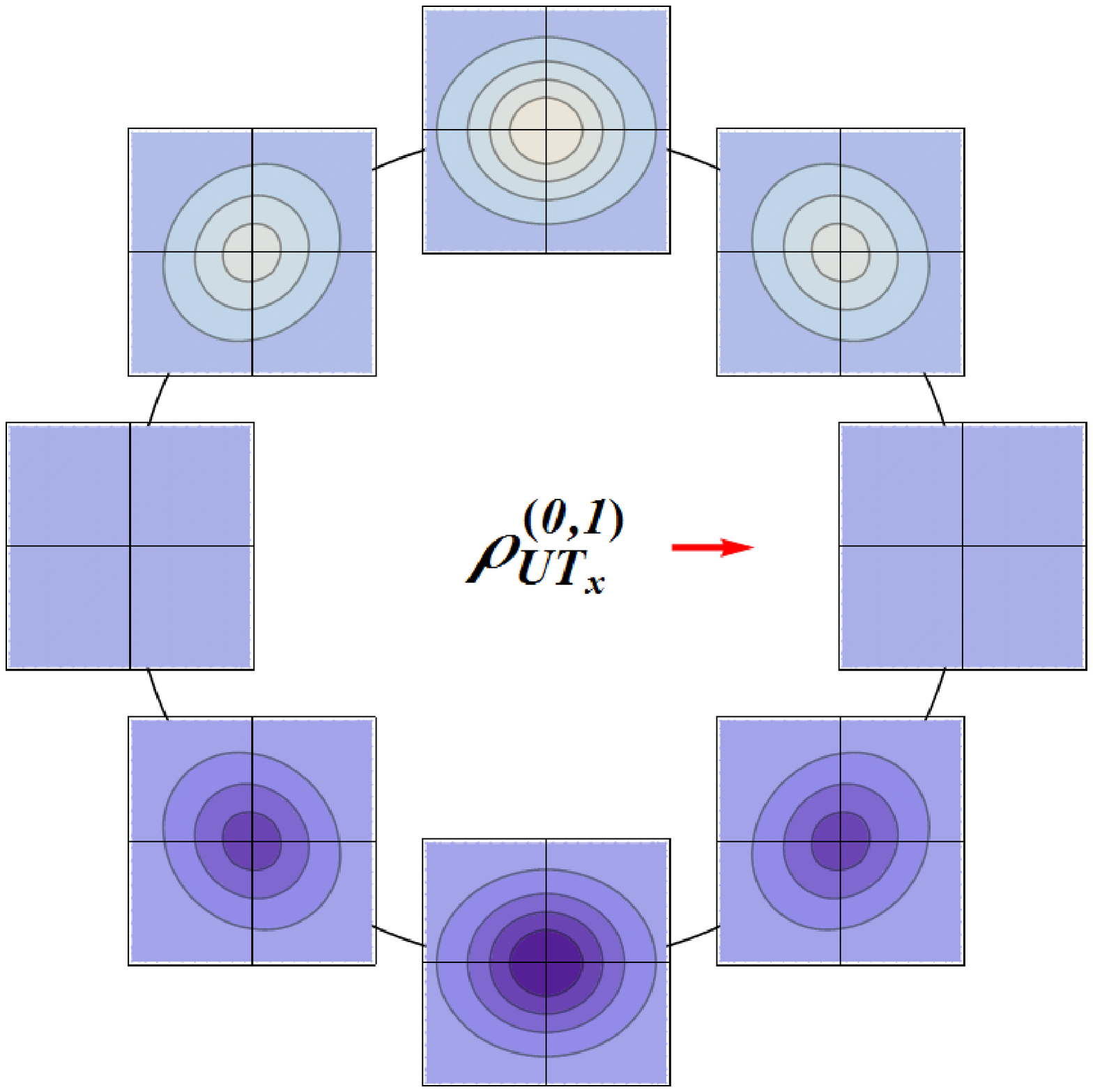}\hspace{1.5cm}\includegraphics[width=7cm]{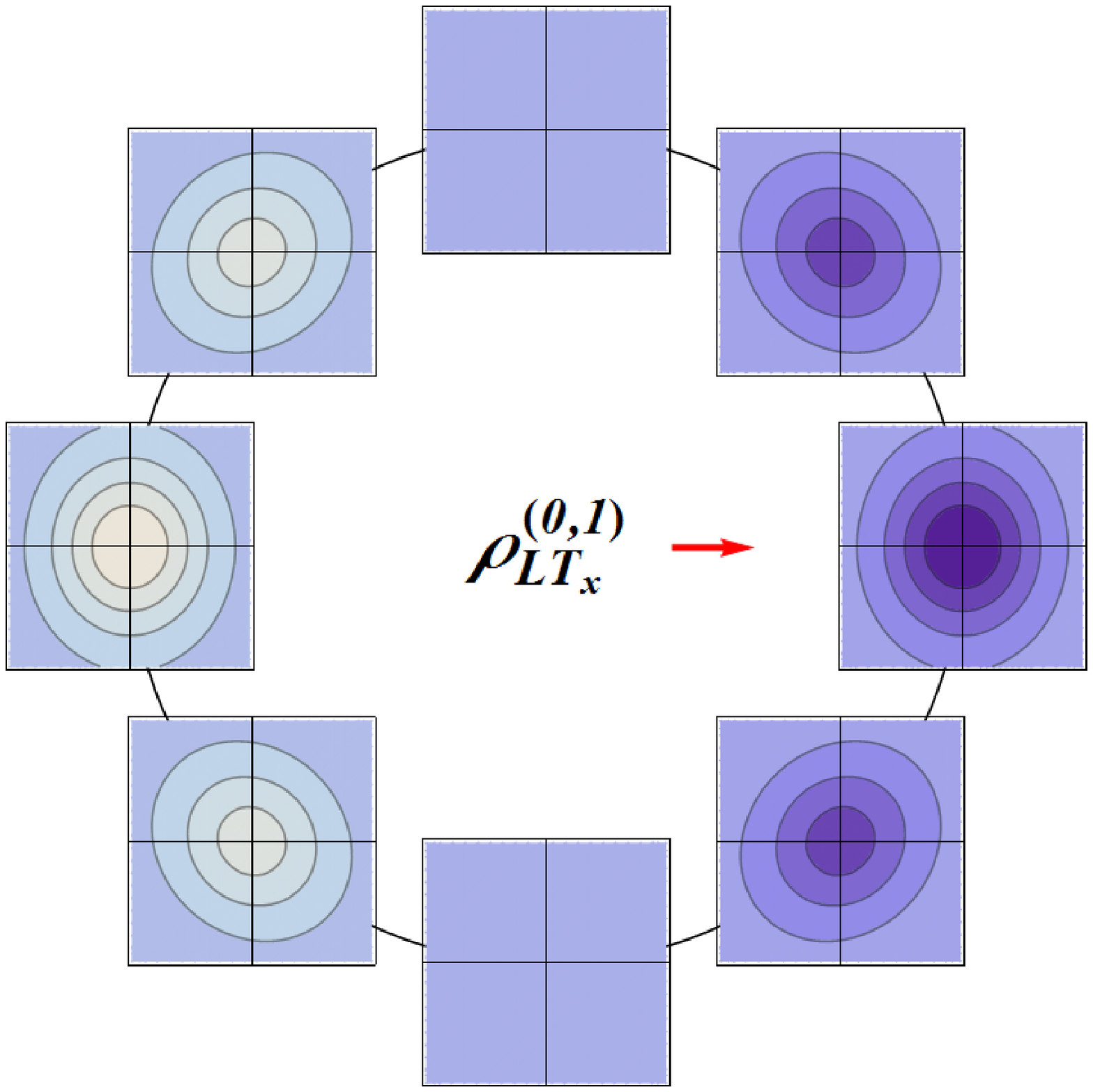}}
\vspace*{8pt}
\caption{The $(0,1)$ multipoles appearing in e.g. $\rho^e_{UT}$ and $\rho^o_{LT}$. The arrow indicates the transverse polarization. See text for more details. \label{fig2}}
\end{figure}

The $(0,1)$ multipoles appear in $\rho^e_X$ with $X=UT,TU$ and in $\rho^o_X$ with $X=LT,TL$, see Fig.~\ref{fig2}. In $\rho^e_X$, the $\uvec b_T$-dipole is orthogonal to the transverse polarization $(\uvec S^q_T\times \hat b_T)_L$, $(\uvec S_T\times \hat b_T)_L$, and generates a shift in impact-parameter space which finds its physical origin in the intrinsic correlation between the transverse polarization and the quark OAM~\cite{Burkardt:2005hp}. In $\rho^o_X$, the $\uvec b_T$-dipole is parallel to the transverse polarization $S_L(\uvec S^q_T\cdot\hat b_T)$, $S^q_L(\uvec S_T\cdot\hat b_T)$, and is associated with an extrinsic double (longitudinal-transverse) spin-orbit correlation. These dipole structures do not survive integration over $\uvec b_T$ and cannot therefore be accessed with TMDs. $\rho^e_X$ however survive integration over $\uvec k_T$ and are then related to the GPDs $(2\tilde H_T+E_T,E)$, respectively.

\begin{figure}[t!]
\centerline{\includegraphics[width=7cm]{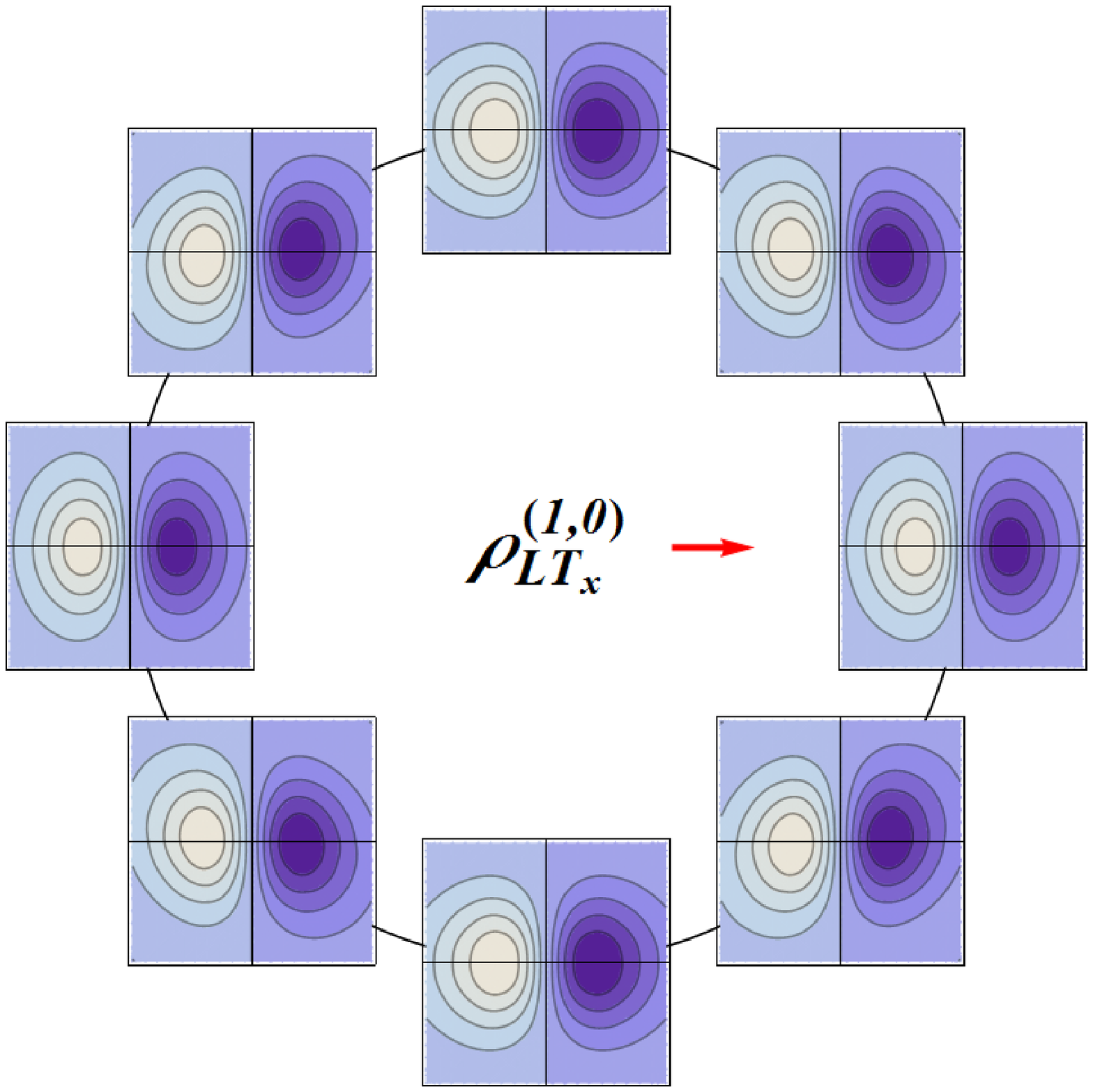}\hspace{1.5cm}\includegraphics[width=7cm]{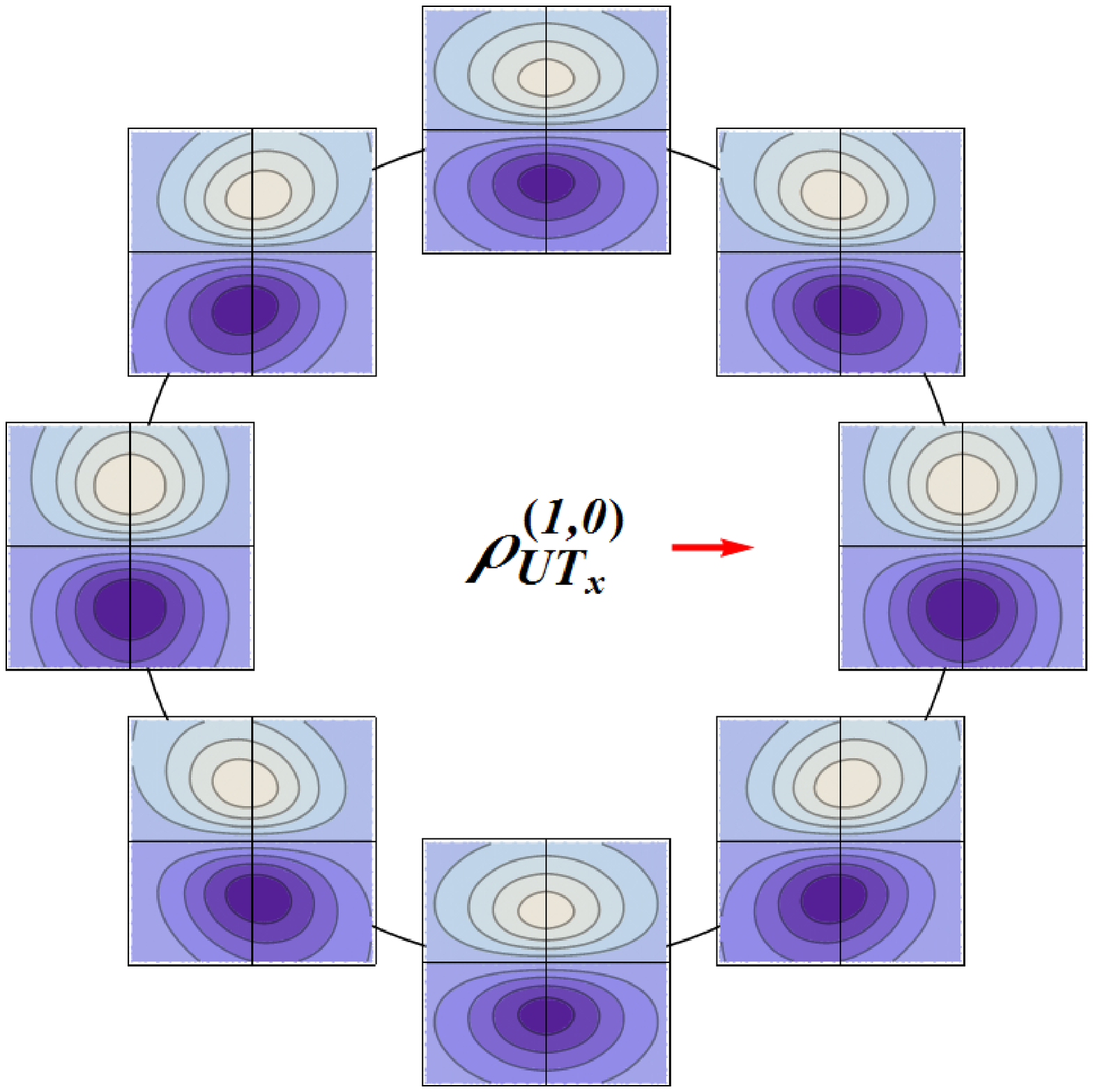}}
\vspace*{8pt}
\caption{The $(1,0)$ multipoles appearing in e.g. $\rho^e_{LT}$ and $\rho^o_{UT}$. The arrow indicates the transverse polarization. See text for more details. \label{fig3}}
\end{figure}

Similarly, the $(1,0)$ multipoles appear in $\rho^e_X$ with $X=LT,TL$ and in $\rho^o_X$ with $X=UT,TU$, see Fig.~\ref{fig3}. In $\rho^e_X$, the $\uvec k_T$-dipole is parallel to the transverse polarization $S_L(\uvec S^q_T\cdot\hat k_T)$, $S^q_L(\uvec S_T\cdot\hat k_T)$, and generates a shift in momentum space which is associated with an intrinsic double (longitudinal-transverse) spin-orbit correlation. In $\rho^o_X$, the $\uvec k_T$-dipole is orthogonal to the transverse polarization $(\uvec S^q_T\times\hat k_T)_L$, $(\uvec S_T\times\hat k_T)_L$, and indicates the presence of a net transverse flow originating from an extrinsic correlation between the transverse polarization and the quark OAM, reminiscent of a quantum Hall effect. These dipole structures do not survive integration over $\uvec k_T$ and cannot therefore be accessed with GPDs. $\rho^e_X$ and $\rho^o_X$ however survive integration over $\uvec b_T$ and are then related to the TMDs $(h^\perp_{1L},g_{1T})$ and $(h^\perp,f_{1T}^\perp)$, respectively.

\subsection{$(1,1)$ multipoles}

\begin{figure}[t!]
\centerline{\includegraphics[width=7cm]{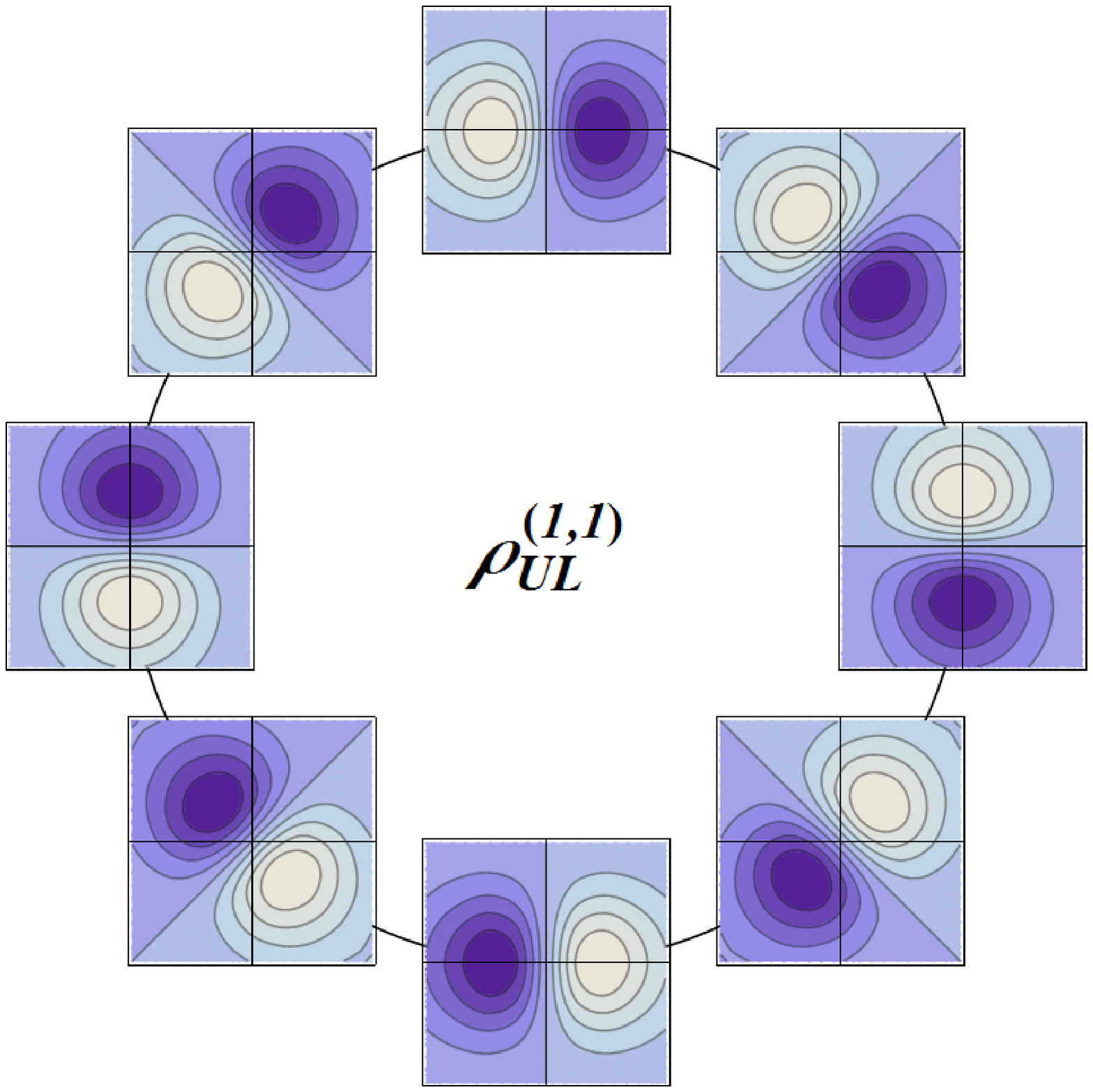}\hspace{1.5cm}\includegraphics[width=7cm]{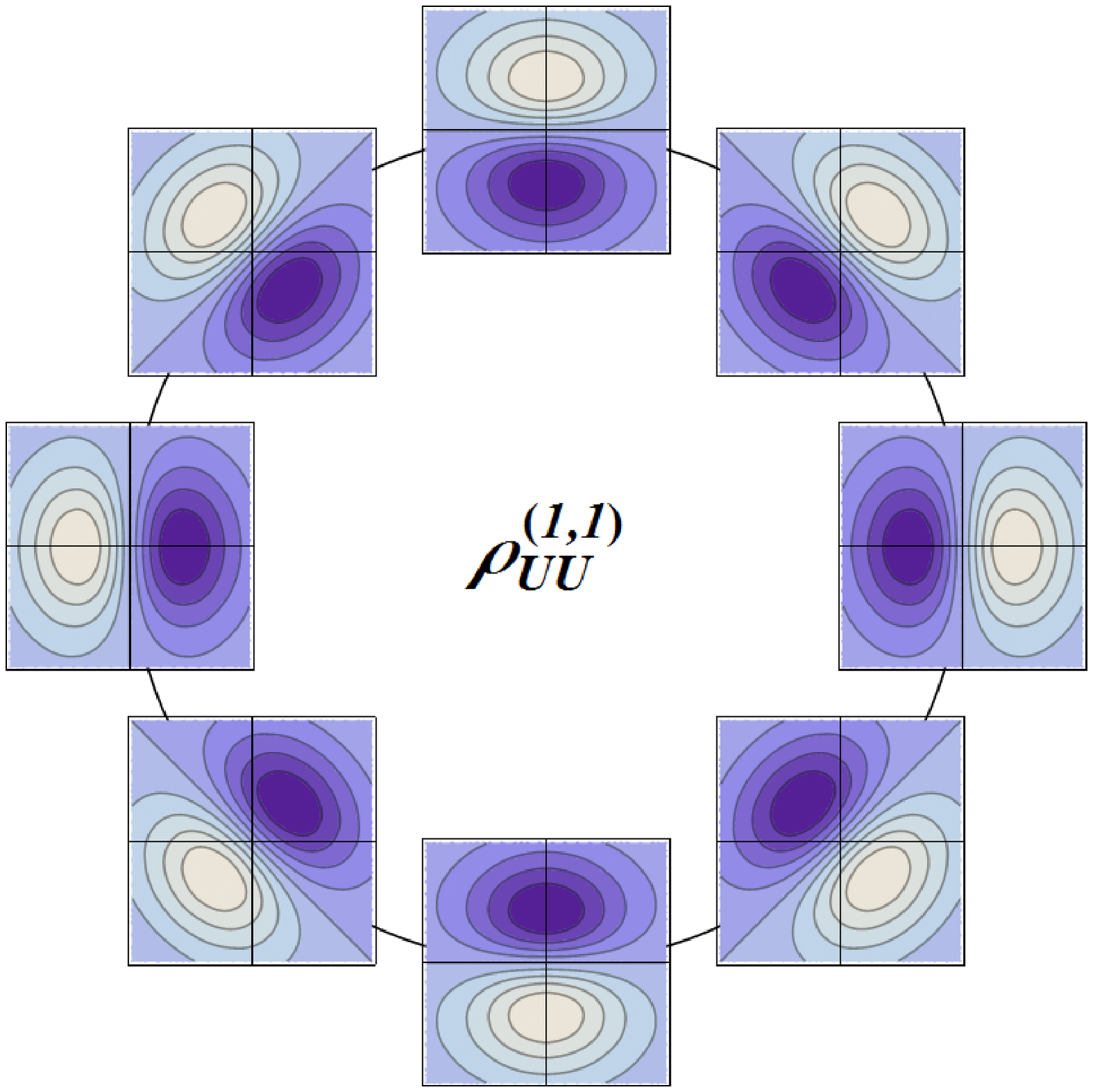}}
\vspace*{8pt}
\caption{The $(1,1)$ multipoles appearing in e.g. $\rho^e_{UL}$ and $\rho^o_{UU}$. See text for more details. \label{fig4}}
\end{figure}

The last multipoles we will discuss here are the ones with $m_k=m_b=1$. Since $m_k=m_b$, these multipoles are invariant under rotation about the longitudinal direction. They appear in $\rho^e_X$ with $X=UL,LU$ and $\rho^o_X$ with $X=UU,LL,TT,TT'$, see Fig.~\ref{fig4}. In $\rho^e_X$ and $\rho^o_{TT'}$, the $\uvec k_T$-dipole is oriented along the polar direction $S^q_L(\hat b_T\times\hat k_T)_L$, $S_L(\hat b_T\times\hat k_T)_L$ and $(\uvec S_T\times\uvec S^q_T)_L(\hat b_T\times\hat k_T)_L$. Clearly, $\rho^e_X$ is related to the orbital motion of quarks correlated with the longitudinal polarization~\cite{Lorce:2011kd,Hatta:2011ku,Lorce:2011ni,Kanazawa:2014nha,Lorce:2014mxa,Liu:2015xha}. $\rho^o_{TT'}$ is less transparent as it originates from an extrinsic double (transverse-transverse) spin-orbit correlation. In $\rho^o_X$ with $X=UU,LL,TT$, the $\uvec k_T$-dipole is oriented along the radial direction $(\hat b_T\cdot\hat k_T)$, $S_LS^q_L(\hat b_T\cdot\hat k_T)$ and $(\uvec S_T\cdot\uvec S^q_T)(\hat b_T\cdot\hat k_T)$, and indicates a net expansion or contraction of the target in the transverse plane as a consequence of the initial and/or final-state interactions, the most simple manifestation of the lensing effect in QCD. Unfortunately, none of these structures survive integration over $\uvec k_T$ or $\uvec b_T$ and cannot therefore be accessed with GPDs or TMDs.

\section{Conclusions}
\label{section:5}

We presented and discussed a selection of leading-twist quark Wigner distributions in the nucleon, introducing a multipole analysis in the transverse phase space. 
In this approach, the multipole structures are constrained by parity and time-reversal symmetries and are multiplied by coefficient functions which depend on $\mathsf P$ and $\mathsf T$-invariant variables only. This representation has several advantages: it provides a clear interpretation of all the amplitudes in terms of the possible correlations between target and quark angular momenta in the transverse phase space, and it provides a convenient basis to make a direct connection with GPDs in impact-parameter space and TMDs in transverse-momentum space. A new graphical representation has also been proposed to display these transverse phase-space distributions.

We presented results for a few lowest multipole structures in both impact-parameter and transverse momentum spaces, half of them being naive $\mathsf T$-even and representing the intrinsic structure of the target, the other half being  naive $\mathsf T$-odd and representing the response of the target to extrinsic initial and/or final-state interactions. These structures have been confirmed and calculated within a light-front consituent quark model

\begin{acknowledgments}

For a part of this work has been supported by the Belgian Fund F.R.S.-FNRS \emph{via} the contract of Charg\'e de recherches.
\end{acknowledgments}

\end{document}